\documentstyle[preprint,aps,prd,psfig]{revtex}

\tightenlines

\newcommand{\AmS}{{\protect\the\textfont2
  A\kern-.1667em\lower.5ex\hbox{M}\kern-.125emS}}
\newcommand{\eqn}[1]{(\ref{#1})}

\newcommand{\vev}[1]{\left\langle #1 \right\rangle}
\newcommand{\be}{\begin{equation}}
\newcommand{\ee}{\end{equation}}
\newcommand{\ben}{\begin{eqnarray}}
\newcommand{\een}{\end{eqnarray}}
\newcommand{\nn}{\nonumber}
\def\lsim{\raise0.3ex\hbox{$<$\kern-0.75em\raise-1.1ex\hbox{$\sim$}}}
\def\gsim{\raise0.3ex\hbox{$>$\kern-0.75em\raise-1.1ex\hbox{$\sim$}}}
\def\simgt{\rlap{\lower 3.5 pt\hbox{$\mathchar \sim$}}\raise 1pt \hbox {$>$}}
\def\simlt{\rlap{\lower 3.5 pt\hbox{$\mathchar \sim$}}\raise 1pt \hbox {$<$}}
\newcommand{\con}{{\rm con}}

\newcommand{\mf}{{\rm MF}}
\newcommand{\msbar}{{\overline {\rm MS}}}
\newcommand{\spr}{{s^\prime}}

\begin{document}
\draft

\title{
\vspace{-5.5cm}
\begin{flushright}  
{\normalsize UTHEP-404}\\
{\normalsize UTCCP-P-66}\\
{\normalsize KANAZAWA-00-03}\\
\end{flushright}
Domain Wall Fermions in Quenched Lattice QCD
}

\author{Sinya Aoki$^a$, Taku Izubuchi$^b$, Yoshinobu Kuramashi$^c$
\thanks{On leave from Institute of Particle and Nuclear Studies,
High Energy Accelerator Research Organization(KEK),
Tsukuba, Ibaraki 305-0801, Japan}
and Yusuke Taniguchi$^a$}

\address{$^a$Institute of Physics, University of Tsukuba, 
Tsukuba, Ibaraki 305-8571, Japan \\
$^b$ Department of Physics, Kanazawa University, 
Kanazawa, Ishikawa 920-1192, Japan \\
$^c$Department of Physics, Washington University, 
St. Louis, Missouri 63130, USA \\
}

\date{\today}

\maketitle

\vspace{-0.5cm}

\begin{abstract}
We study the chiral properties and the validity of perturbation theory 
for domain wall fermions in quenched lattice QCD at $\beta=6.0$. 
The explicit chiral symmetry breaking term in the axial Ward-Takahashi 
identity is found to be very small already at $N_s=10$,
where $N_s$ is the size of the fifth dimension, and
its behavior seems consistent with
an exponential decay in $N_s$ within the limited range of $N_s$ we explore.
From the fact that the critical quark mass, 
at which the pion mass vanishes 
as in the case of the ordinary Wilson-type fermion, 
exists at finite $N_s$,
we point out that this may be a signal
of the parity broken phase and investigate the possible existence of 
such a phase in this model at finite $N_s$.
The $\rho$ and $\pi$ meson decay constants 
obtained from the four-dimensional local currents
with the one-loop renormalization factor
show a good agreement with those obtained from the 
conserved currents.
\end{abstract}


\newpage
\section{Introduction}

Domain wall fermion model\cite{kaplan,shamir,FS} 
is a 5-dimensional Wilson fermion with free boundaries in the fifth dimension.
At each of the boundaries 
either of the left- and right-handed chiral mode 
is expected to be localized exponentially,
and hence the mixing between them, which represents the explicit 
chiral symmetry breaking effects, is suppressed exponentially. 
In the large $N_s$ limit, where $N_s$ is the size of the fifth direction,
domain wall QCD (DWQCD) would have the desirable features, such 
as (i) no fine tuning for the chiral limit,
(ii) $O(a^2)$ scaling violation 
and (iii) no mixing between three- and four-quark operators with
different chiralities.
It is shown that these expectations are realized
within the perturbation theory up 
to the one-loop level in the limit $N_s\rightarrow \infty$
\cite{pt_at,pt_awi,pt_2,pt_34,NT99}. Pioneering numerical studies 
also seem to support these expectations\cite{BS,lat98}.

In this paper we focus on two issues. One is the chiral properties 
of DWQCD at the finite $N_s$.
It is of great importance to understand correctly
how the ideal chiral properties are
realized nonperturbatively as $N_s$ increases. 
We investigate the $N_s$ dependence of the explicit chiral symmetry breaking 
term in the axial Ward-Takahashi identity, 
which are expected to be exponentially suppressed as the $N_s$ increases.
Correspondingly, at finite $N_s$, the pion remains massive in the chiral limit.
We notice, however, that the pion mass at finite $N_s$ may vanish
at some non-zero value of quark mass, which we call the critical quark mass,
in a similar manner that the pion mass in the Wilson fermion formulation
vanishes at the critical hopping parameter.
In the case of the Wilson fermion formulation,
it is well known that the massless pion 
implies the existence of a parity(-flavor) broken phase\cite{WPBP}.
The analysis of two-dimensional Gross-Neveu model with domain-wall 
fermions indeed demonstrates the existence of the parity broken phase 
at the finite $N_s$, and the negative critical quark mass goes to zero
as $N_s$ increases\cite{PBP_v,PBP_in}.
In this paper we explore the negative quark mass region to examine 
the possible existence of the parity broken phase in the four-dimensional 
DWQCD.

Another important issue in this paper is the validity of the
perturbation theory for DWQCD. Recently we calculated the renormalization 
factors for the bilinear operators\cite{pt_2} and three- 
and four-quark operators\cite{pt_34}, for the latter of which we find 
the multiplicative renormalizability at infinite $N_s$ as opposed
to the Wilson fermion case. We test the validity of the
perturbation theory comparing
the $\rho$ and $\pi$ meson decay constants extracted from 
the local currents with those from the conserved ones.

This paper is organized as follows.
In Sec.~\ref{sec:parameters} we present the details of simulation including
the action of domain wall fermion.
The chiral properties of DWQCD are investigated 
in Sec.~\ref{sec:chiral}, where we also examine the existence of
the parity broken phase.
In Sec. \ref{sec:f_meson} we test the validity of the perturbation theory
for DWQCD using the meson decay constants.
Our conclusions and discussions are summarized in 
Sec.~\ref{sec:conclusions}.

\section{Details of Numerical Simulation}
\label{sec:parameters}
\subsection{Action}

The domain wall fermion action is written as\cite{shamir,FS} 
\ben
S_f &=& -\sum_{x,s,y,\spr} {\bar \psi}(x,s) D_{dwf}(x,s;y,\spr) \psi(y,\spr)~,
\label{eq:action}\\
D_{dwf}(x,s;y,\spr) &=& 
(5-M)\delta_{x,y} \delta_{s,\spr} -D^4(x,y)\delta_{s,\spr} 
-D^5(s,\spr)\delta_{x,y}~,\label{eq:qm_dwf}\\
D^4(x,y) &=& \frac{1}{2}\left[(1-\gamma_\mu)U_{x,\mu}\delta_{x+\hat\mu,y}
          + (1+\gamma_\mu)U^\dagger_{y,\mu}\delta_{x-\hat\mu,y}\right],\\
D^5(s,\spr) &=& \left\{
\begin{array}{lll}
P_L\delta_{2,\spr} &\ & (s=1)\\
P_L\delta_{s+1,\spr} & +P_R\delta_{s-1,\spr} & (1<s<N_s)\\
&P_R\delta_{N_s-1,\spr} & (s=N_s)
\end{array}
\right.
\een
with domain wall height $M$ and the extent of the fifth dimension $N_s$.
The right- and left-handed projection operators are defined by
$P_{R,L}=(1\pm\gamma_5)/2$. 
The chiral zero mode is supposed to appear for $M>0$. 
The index $x,y$ stands for four-dimensional lattice sites, while
the index $s,\spr$ for the fifth direction runs from 1 to $N_s$. 

The quark fields on the four-dimensional space-time 
are given by combinations of the fermion fields at the boundaries,
\ben
q(x)= P_L\psi(x,1) +P_R\psi(x,N_s)~,
\label{eq:q}\\
\bar q(x)= \bar\psi(x,1)P_R +\bar\psi(x,N_s)P_L~,
\label{eq:qbar}
\een
where the left(right) chiral zero mode is well localized at $s= 1(N_s)$
for $ 0 < M < 2$ in the free theory.

The explicit bare quark mass $m_f$ is introduced as
\begin{equation}
S_f \rightarrow S_f - \sum_x m_f \bar q(x)  q(x)~,
\end{equation}
which provides the quarks in eqs.(\ref{eq:q}) and (\ref{eq:qbar}) 
with a Dirac mass ${\bar m_f}$.
For the free quark case, we obtain ${\bar m_f}=M(2-M)m_f$
in the $N_s\longrightarrow \infty$ limit\cite{shamir}.

The action (\ref{eq:action}) is invariant under global SU($N_f$) symmetry,
which yields the five-dimensional conserved current\cite{FS},
\ben 
j^a_\mu(x,s)&=&\frac{1}{2}\left[{\bar \psi}(x,s)(-1+\gamma_\mu)U_{x,\mu}
\frac{\lambda^a}{2}\psi(x+{\hat \mu},s)
+{\bar \psi}(x+{\hat \mu},s)(1+\gamma_\mu)U^\dagger_{x,\mu}
\frac{\lambda^a}{2}\psi(x,s)\right],
\een
where $\lambda^a/2$ $(a=1,\dots,N_f^2-1)$ are generators of SU($N_f$) group.
In terms of this current we can define the conserved
vector and axial currents,
\ben
V_\mu^{a,\con}(x)&=&\sum_{s=1}^{N_s} j_\mu^a(x,s), 
\label{eq:vcon}\\
A_\mu^{a,\con}(x)&=&-\sum_{s=1}^{N_s}{\rm sign}
\left(\frac{N_s}{2}-s+\frac{1}{2}\right) j_\mu^a(x,s). 
\label{eq:acon}
\een
We should note that $A_\mu^{a,\con}$ is not exactly conserved in
the finite $N_s$ contrary to the vector case: the divergence 
of $A_\mu^{a,\con}$ satisfies
\begin{eqnarray}
\vev{\nabla_\mu^- A_\mu^{a,\con}(x) {\cal O}(y)}
=2m_f \vev{{\bar q}(x)\gamma_5\frac{\lambda^a}{2}q(x) {\cal O}(y)}
+2\vev{J^a_{5q}(x) {\cal O}(y)}
\label{eq:pcac}
\end{eqnarray}
with
\be
J^a_{5q}(x)=-{\bar \psi}(x,N_s/2)P_L\frac{\lambda^a}{2}\psi(x,N_s/2+1)
+{\bar \psi}(x,N_s/2+1)P_R\frac{\lambda^a}{2}\psi(x,N_s/2),
\ee
where $\nabla_\mu^-$ is the backward difference operator:
$\nabla_\mu^- f(x) = f(x) - f(x-\mu )$
and ${\cal O}$ is some operator.
We will set $x \neq y$ in our numerical simulation.
The contribution of $J^a_{5q}(x)$ in eq.(\ref{eq:pcac}) 
is expected to be exponentially suppressed as the $N_s$ increases
for the physical operator ${\cal O}_{\rm phys.}(\bar q,q)$.

\subsection{Simulation parameters}

The simulation is carried out with the plaquette gauge action at 
$\beta=6.0$ on $16^3\times 32\times N_s$ lattices in quenched QCD. 
Table~\ref{tab:parameter} summarizes our run parameters.
We generated gauge configurations on a $16^3\times 32$ lattice 
with the five-hit pseudo heat-bath algorithm. Each configuration is
separated by 2000 sweeps after 20000 sweeps for the initial
thermalization. In DWQCD the same gauge configuration is assigned on each
layer of the fifth direction. The $\rho$ meson mass 
at the chiral limit ($m_f = 0$)
is used to determine the lattice spacing, which gives $a^{-1}\sim 2$GeV
with $m_\rho=770$MeV as input while slightly depending 
on $M$ and $N_s$ (see Sec.~\ref{sec:chiral}).
The mean-field estimate for the optimal value of $M$ gives\cite{pt_at}
\be
M=1+4(1-u)=1.819~, \label{eq:MFanz}
\ee
employing $u=1/(8 K_c)$ with $K_c=0.1572$ 
for the Wilson fermion at $\beta =6.0$.

The quark propagator is obtained by solving the inverse of 
the quark matrix (\ref{eq:qm_dwf}) with the unit wall source 
without gauge fixing,
\ben
&&\sum_{y,\spr} \sum_{b} D^{ab}_{dwf}(x,s;y,\spr) G^{b}_\psi(y,\spr) 
= B^{a}(x,s)~,
\label{eq:qprop}\\
&&~~~~B^{a}(x,s) = {1\over 3 V_s} \sum_c \delta_{ac}
\delta_{t,0}\sum_{\vec x^\prime}\delta_{{\vec x},{\vec x^\prime}}
\left[\delta_{s,1}P_R + \delta_{s,N_s}P_L\right]~,
\een
where $x=({\vec x},t)$ with spatial coordinates $\vec x$ and time coordinate
$t$; $V_s$ is the spatial volume
and $a,b,c$ denote the color indices. 
The four-dimensional quark propagator is constructed with 
a linear combination of $G_\psi$,
\be
G_q(x) = P_L G_\psi(x,1) + P_R G_\psi(x,N_s)~.
\ee
To solve the linear equation (\ref{eq:qprop}),
we employ the conjugate gradient 
method with even/odd preconditioning\cite{evenodd}. 
In the case of $m_f=0.025$, 
about 500 iterations are needed to 
satisfy a convergence criterion
which requires $|r|^2/|G_\psi|^2<10^{-12}$ with the residual $r$.
Neither minimal residue(MR) nor BiCGStab 
converges within a practical number of iterations for 
gauge configurations at $\beta\sim6.0$. 
Throughout this paper, statistical errors are 
estimated by the single elimination jackknife method.

\section{Chiral properties of DWQCD}
\label{sec:chiral}
\subsection{Restoration of chiral symmetry}

We investigate the existence of the chiral zero mode expected
in the limit of $m_f\to 0$ and $N_s\to\infty$ using the
pion mass at the chiral limit ($m_f =0$)
and the anomalous contribution $J_{5q}$ in the PCAC relation (\ref{eq:pcac}).
The pion mass $m_{\pi}(m_f,N_s,M)$ is extracted 
from a global fit of the two-point function
\be
\sum_{{\vec x},{\vec y}}\vev{J_\pi({\vec x},t) J_\pi^\dagger({\vec y},0)}
\ee
employing the fitting function
\be
V_s\frac{|\vev{0|J_\pi|\pi}|^2}{2m_\pi}
\left({\rm e}^{-m_\pi t}+{\rm e}^{-m_\pi(32-t)}\right), 
\ee
where $J_\pi(x)={\bar q}(x)\gamma_5 q(x)$.
We choose the fitting range to be $9\le t\le 23$ after taking account
of the time reversal symmetry $t\rightarrow 32-t$.

\begin{figure}[t]
\hspace{10pt}\psfig{file=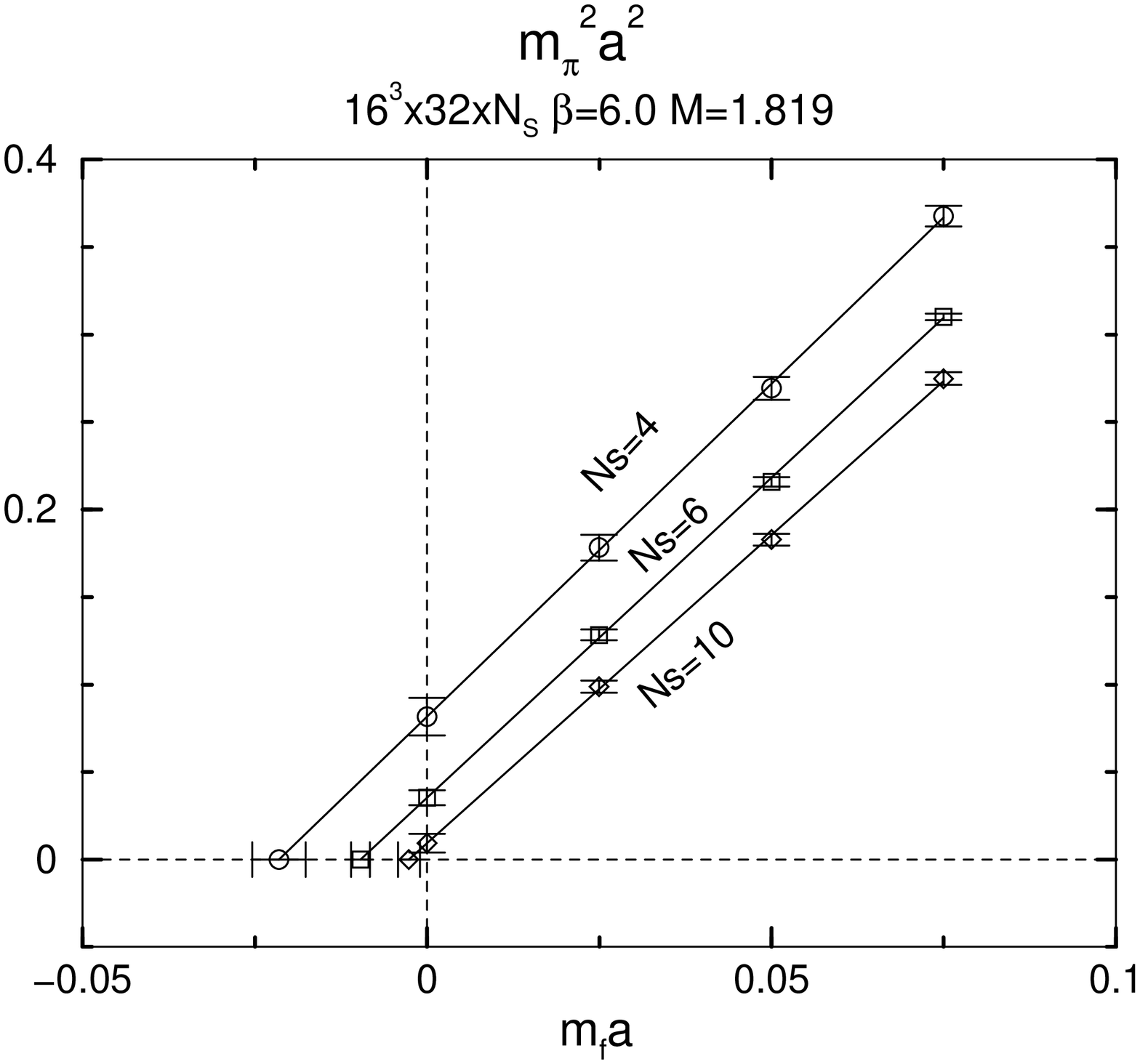,width=100mm,angle=-90}
\caption{Pion mass squared as a function of $m_f a$ 
at $M=1.819$ and $ N_s=4, 6, 10$. Solid lines show linear fits.
}
  \label{fig:Mpi2_mf}
\end{figure}

\begin{figure}[t]
\hspace{10pt}\psfig{file=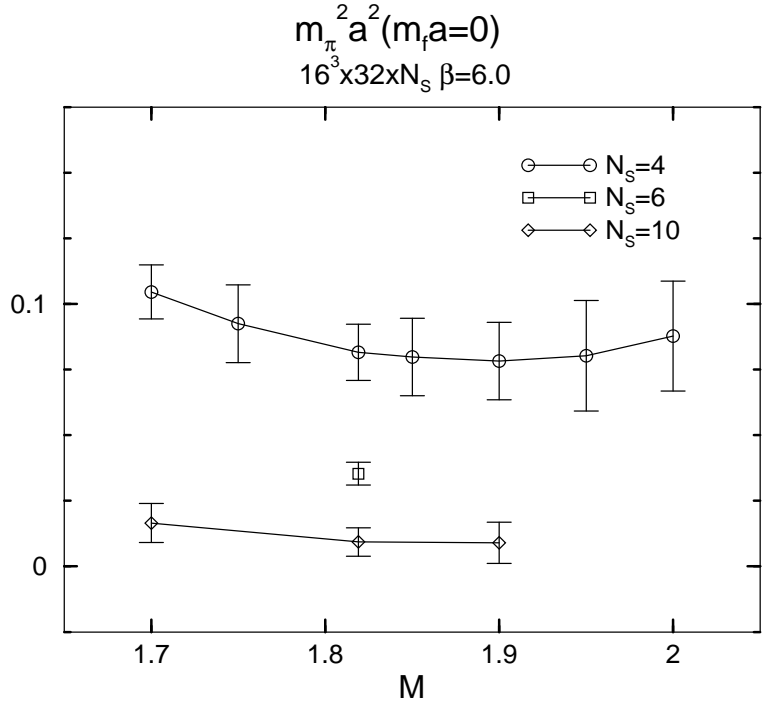,width=100mm,angle=-90}
\caption{Pion mass squared extrapolated to $m_f\to 0$ as a function of $M$ 
at $ N_s=4, 6, 10$. }
  \label{fig:Mpi2_M}
\end{figure}

\begin{figure}[t]
\hspace{10pt}\psfig{file=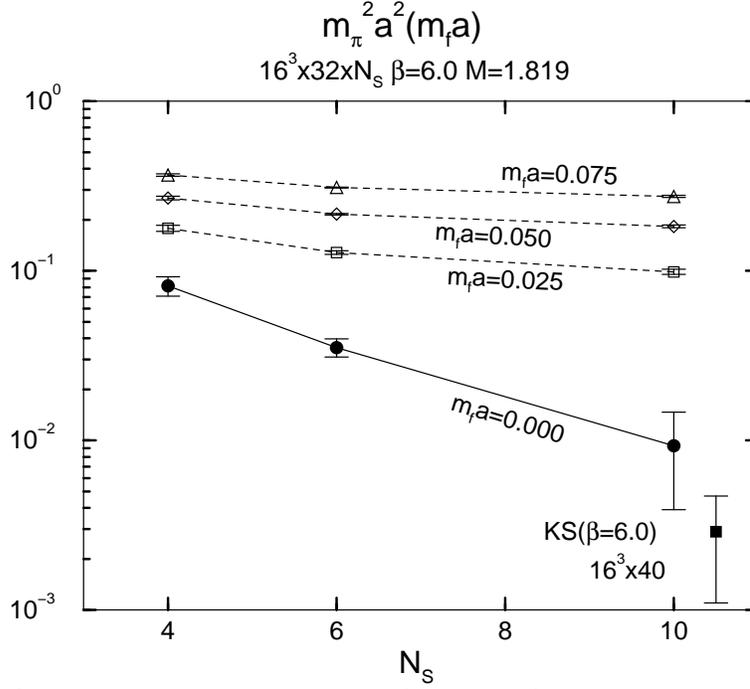,width=100mm,angle=-90}
\caption{Pion mass squared as a function of $N_s$ 
at $M=1.819$. 
Filled square shows the value for the Nambu-Goldstone pion of
the Kogut-Susskind fermion at $\beta=6.0$ on a $16^3\times40$ lattice.
}
  \label{fig:Mpi2_Ns}
\end{figure}

Figure~\ref{fig:Mpi2_mf} shows the $m_fa$ dependence of $m_\pi^2$.
We observe a good linear behavior for the pion mass squared 
as a function of $m_f$. Employing a linear extrapolation, 
we obtain the value of $m_\pi^2$ at $m_f=0$, which
is shown in Fig.~\ref{fig:Mpi2_M} as a function of $M$.
We find that the minimum point of the pion mass is around $M=1.819$, 
which shows the mean-field estimation of optimal $M$ 
in (\ref{eq:MFanz}) is reasonably accurate.
Furthermore, extending the linear extrapolation into the negative
$m_f$ region, as is done for the ordinary Wilson-type fermion,
we can also extract the critical quark mass $m_c(N_s,M)$ 
at which the pion mass squared vanishes. 
As we expected, both $m_\pi^2$ at $m_f=0$ and  $m_c(N_s,M)$ 
decreases monotonically as $N_s$ increases.
In Fig.~\ref{fig:Mpi2_Ns} we plot $m_\pi^2$ 
for $m_f a =$0.075, 0.050, 0.025 and 0 as a
function of $N_s$, where $m_\pi^2(m_f=0)$ is the extrapolated value.
The magnitude of $m_\pi^2$ at $m_f=0$ seems to diminish
exponentially in $N_s$.
In order to confirm that $m_\pi^2$ at $m_f=0$ indeed vanishes exponentially
in $N_s$, one has to further increase $N_s$.
However, we do not attempt such an investigation at the present lattice size,
since $m_\pi^2(m_f=0)$ at $N_s=10$ is already as small as that 
for the Nambu-Goldstone pion 
of the Kogut-Susskind fermion, which gives an estimate of the finite size 
effect
to the would-be massless pion at the same $\beta$ and the spatial lattice size.

It is reported, at $\beta = 6.0$ on a $16^3\times 32$ lattice,
that $m_\pi^2 = 0.014(2)$ in the chiral limit
for $N_s=16$ and $M=1.8$\cite{RBC},
which is roughly equal to our value of 
$m_\pi^2 =0.0093(54)$ for $N_s =10$ and $M=1.819$.

The anomalous contribution due to $J_{5q}$ in eq.(\ref{eq:pcac})
is evaluated by calculating the ratio of two-point functions,
\be
\frac{\sum_{{\vec x},{\vec y}}\vev{J_{5q}({\vec x},t) 
J_\pi^\dagger({\vec y},0)}}
{\sum_{{\vec x},{\vec y}}\vev{J_\pi({\vec x},t) J_\pi^\dagger({\vec y},0)}}
\stackrel{t\gg0}{\longrightarrow} 
\frac{\vev{0|J_{5q}|\pi}}{\vev{0|J_\pi|\pi}},
\ee
where we employ a constant fit over the range $8\le t\le 24$. 
This quantity measures contribution 
of the explicit chiral symmetry breaking effects in the PCAC relation:
\be
\frac{\vev{0|2J_{5q}|\pi}}{\vev{0|J_\pi|\pi}}
=\frac{\vev{0|\nabla_\mu^- A_\mu^\con|\pi}}{\vev{0|J_\pi|\pi}}-2m_f.
\ee

\begin{figure}[t]
\hspace{10pt}\psfig{file=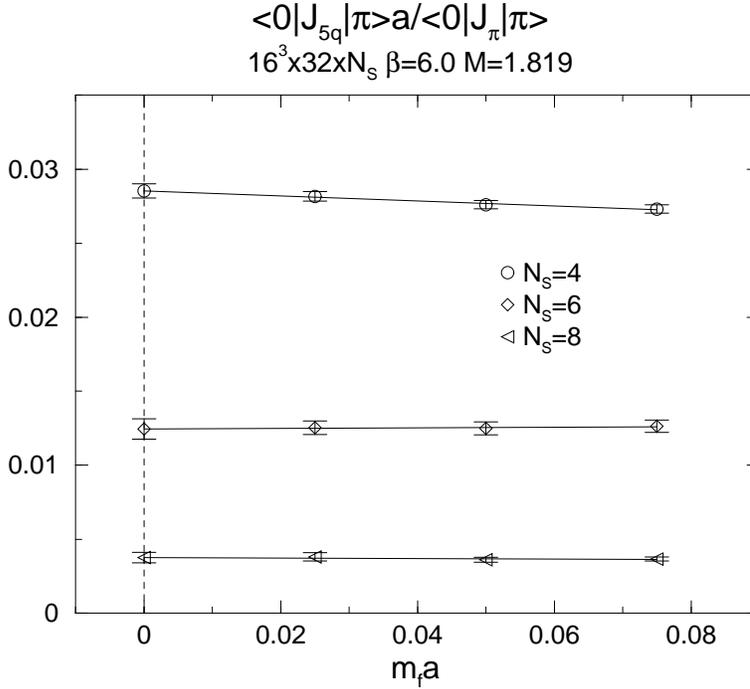,width=100mm,angle=-90}
\caption{${\vev{0|J_{5q}|\pi}}/{\vev{0|J_\pi|\pi}}$
extrapolated as a function of  $m_fa$ at $M=1.819$ with $N_s=4,6,8$.
Solid lines show linear fits.}
  \label{fig:J5q_mf}
\end{figure}

\begin{figure}[t]
\hspace{10pt}\psfig{file=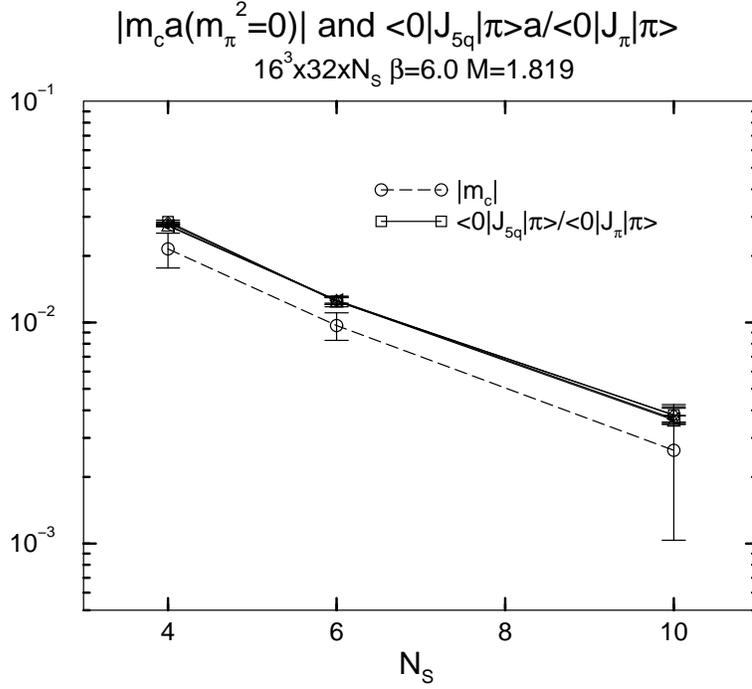,width=100mm,angle=-90}
\caption{${\vev{0|J_{5q}|\pi}}/{\vev{0|J_\pi|\pi}}$
as a function of  $N_s$ at $M=1.819$ for $m_fa=0.075, 0.050, 0.025, 0$
together with the critical quark mass $|m_c|$.}
  \label{fig:ItcX_Ns}
\end{figure}

In Fig.~\ref{fig:J5q_mf} we plot the results 
for ${\vev{0|J_{5q}|\pi}}/{\vev{0|J_\pi|\pi}}$ as a function of $m_fa$. 
The data shows little $m_f$ dependence at each $N_s$.  
We obtain the value of ${\vev{0|J_{5q}|\pi}}/{\vev{0|J_\pi|\pi}}$ at $m_f=0$
by extrapolating the data linearly to $m_f=0$. 
Figure~\ref{fig:ItcX_Ns} illustrates the $N_s$ dependence
of ${\vev{0|J_{5q}|\pi}}/{\vev{0|J_\pi|\pi}}$ at 
$m_f a =$0.075, 0.050, 0.025, 0, 
where we also plot the results for $|m_c|$ for comparison. 
We find that the magnitude of
${\vev{0|J_{5q}|\pi}}/{\vev{0|J_\pi|\pi}}$ decreases exponentially as $N_s$
increases. The same situation is observed for $|m_c|$.
These features suggest that the chiral symmetry would be 
restored in the limit of $m_f\to 0$ and $ N_s\to\infty$ at $\beta=6.0$.

Note that $|m_c|$ and ${\vev{0|J_{5q}|\pi}}/{\vev{0|J_\pi|\pi}}$ are
expected to be consistent, if the latter has no $m_f$ dependence
as seen in Fig.~\ref{fig:J5q_mf}.
However we observe a slight deviation between them in Fig.~\ref{fig:ItcX_Ns}.
We point out two possibilities to yield this inconsistency.
One is the finite size effects on the pion mass and 
the other is the quenched chiral
logarithm plaguing the pion mass in the quenched approximation. 
We should note  that the quantity 
${\vev{0|J_{5q}|\pi}}/{\vev{0|J_\pi|\pi}}$ is
free from both systematic contaminations, which assures that 
${\vev{0|J_{5q}|\pi}}/{\vev{0|J_\pi|\pi}}$ is superior to $|m_c|$
to measure the remnant of the chiral symmetry breaking effects in DWQCD. 

\begin{figure}[t]
\hspace{10pt}\psfig{file=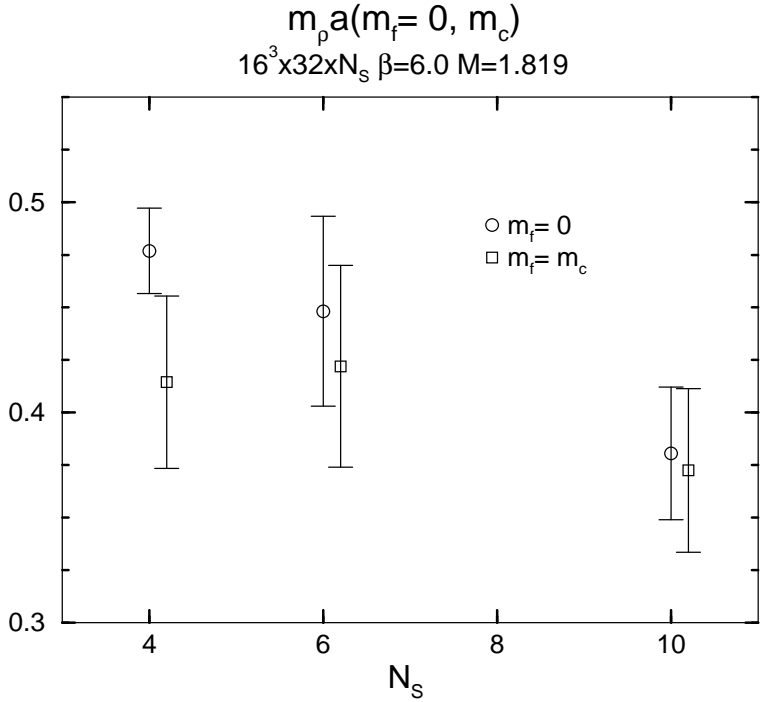,width=100mm,angle=-90}
\caption{$\rho$ meson mass extrapolated  to $m_f=0$ and $m_f=m_c$
as a function of $N_s$ at $M=1.819$ . 
}
  \label{fig:Mrho_Ns}
\end{figure}

We briefly mention the determination of the lattice spacing $a$
from the $\rho$ meson mass.
Figure~\ref{fig:Mrho_Ns} shows the $N_s$ dependence
of $m_\rho a$ at $m_f=0$ (circles) and at $m_f=m_c$ (squares) with
$M=1.819$, which
are obtained by a linear extrapolation from the data.
For small $N_s(=4)$, $m_\rho a$ at $m_f=0$ differs from
the one at $m_f=m_c$ beyond statistical errors, 
while the difference almost vanishes at $N_s=10$.
This is true for other values of $M$:
within statistical errors the values of $m_\rho a$ at $m_f=0$ and $m_f=m_c$ 
are consistent with each other for $M=1.7-1.9$ at $N_s=10$.
Furthermore the $M$ dependence of $m_\rho$ itself is mild over
this range of $M$ at $N_s=10$.
>From the value of $m_\rho a$ at $m_f=0$ with $M=1.819$ and $N_s=10$, we
estimate $a^{-1}=2.02(17)$GeV from $m_\rho=770$MeV as an input.

\subsection{Parity broken phase}

In the Wilson fermion formalism, 
the critical quark mass 
is identified with the second order phase transition point between 
the (spontaneously) parity broken phase and the parity 
symmetric phase\cite{WPBP}. 
The massless pion is understood as the massless particle associated with
the second order phase transition.
Recently, the study of
the two-dimensional Gross-Neveu model with domain wall fermions
pointed out that this picture is true even for this model
at finite $N_s$\cite{PBP_v,PBP_in}.
In this section we examine 
whether the massless pion at $m_f = m_c$ at finite $N_s$
can be understood in this way.

\begin{figure}[t]
\hspace{10pt}\psfig{file=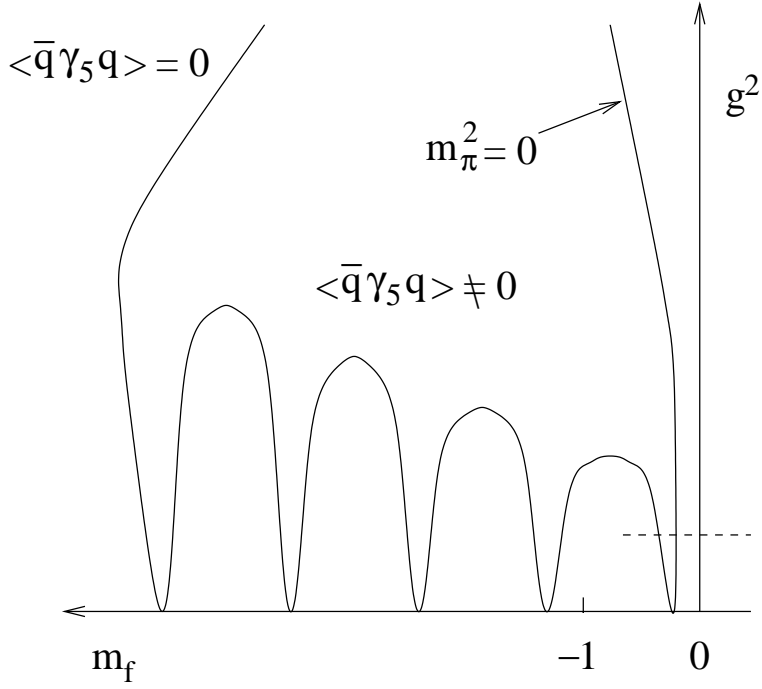,width=100mm,angle=-90}
\caption{Schematic phase diagram in ($g^2,m_f$) plane for $N_s=$even.
See text for dashed segment.}
  \label{fig:pbpdgm}
\end{figure}

For the finite $N_s$, the parity broken phase may exist 
in negative $m_f$ regions. From the study of 
the two-dimensional Gross-Neveu model with domain wall fermions\cite{PBP_in},
we expect the phase diagram in Fig.~\ref{fig:pbpdgm} 
for four-dimensional DWQCD. Each of five critical points, where the
$m_\pi^2=0$ line is touched on $g^2=0$ line at  
$m_f=-(1-M+2l)^{N_s}$ ($l=0,1,2,3,4$), corresponds to the massless
particle pole of momentum
\ben
p_\mu(l=0)&=&(0,0,0,0), \nn\\
p_\mu(l=1)&=&(\frac{\pi}{a},0,0,0),(0,\frac{\pi}{a},0,0)
            ,(0,0,\frac{\pi}{a},0),(0,0,0,\frac{\pi}{a}),  \nn\\
p_\mu(l=2)&=&(\frac{\pi}{a},\frac{\pi}{a},0,0)
            ,(\frac{\pi}{a},0,\frac{\pi}{a},0)
            ,(\frac{\pi}{a},0,0,\frac{\pi}{a})  \\
          &&,(0,\frac{\pi}{a},\frac{\pi}{a},0)
            ,(0,\frac{\pi}{a},0,\frac{\pi}{a})
            ,(0,0,\frac{\pi}{a},\frac{\pi}{a})  \nn\\
p_\mu(l=3)&=&(\frac{\pi}{a},\frac{\pi}{a},\frac{\pi}{a},0)
            ,(\frac{\pi}{a},\frac{\pi}{a},0,\frac{\pi}{a})
            ,(\frac{\pi}{a},0,\frac{\pi}{a},\frac{\pi}{a})
            ,(0,\frac{\pi}{a},\frac{\pi}{a},\frac{\pi}{a}),  \nn\\
p_\mu(l=4)&=&(\frac{\pi}{a},\frac{\pi}{a},\frac{\pi}{a},\frac{\pi}{a}). \nn
\een 
Since the critical point for the $p_\mu(l=0)$ mode converges to $m_f=0$
for $0 < M < 2$ as $N_s$ increases, 
while other four critical points move rapidly to
$|m_f|=\infty$, we can expect one small ($|m_f|<1$) critical point 
and four large ($|m_f|>1$) critical points at finite $N_s$ in four-dimensional
DWQCD.

\begin{figure}[t]
\hspace{10pt}\psfig{file=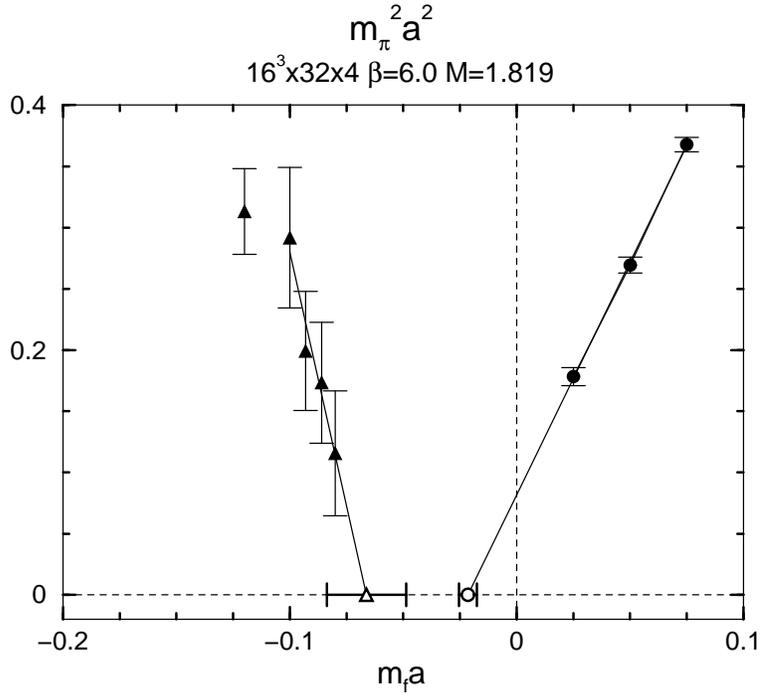,width=100mm,angle=-90}
\caption{$m_\pi^2$ in lattice unit as a function of $m_f a$
at $M=1.819$ and  $N_s=4$. Solid lines show linear extrapolations.
}
  \label{fig:Mpi2_mfNeg}
\end{figure}

Based on these speculation, let us 
investigate existence of the ``cusp''(branch) 
corresponding the $p_\mu(l=0)$ mode, which
provides an evidence of the parity broken phase.
We calculate pion masses at $m_f a = -0.120, -0.100, -0.093, -0.086, -0.080$ 
for $N_s=4$ and $M=1.819$.
Without adding the external field which couples to the parity broken 
order parameter, one can not perform the correct simulation
in the parity broken phase. To avoid the direct simulation in the
parity broken phase, we take rather large magnitudes for negative
$m_f$ values so that the system goes into another symmetric phase,
and try to find another side of the phase boundary (critical quark mass)
from this phase.
In Fig.~\ref{fig:Mpi2_mfNeg} we plotted the $\pi$ meson mass squared 
$m_\pi^2$ as a function of $m_f$. 
Extrapolations of $m_\pi^2$ to zero both 
from positive and negative $m_f$, where 
four largest $m_f$ are used for negative $m_f$, indicate
that a parity broken phase may exist around $m_f a \sim -0.03$.
We can interpret that Fig.~\ref{fig:Mpi2_mfNeg} 
represents the portion in Fig.~\ref{fig:pbpdgm} denoted by dashed segment.
It is noted that the large errors for negative $m_f$
are caused by the fact that the pion propagators form peculiar shapes 
similar to ``W'' character, which has been often observed near the another 
side of the parity broken phase for the Wilson fermion\cite{AKU}.

\section{Meson decay constants}
\label{sec:f_meson}
\subsection{Perturbative renormalization factors}

The $\rho$ meson decay constant $f_\rho$ is defined by
\begin{equation}
\vev{0|V_\mu|\rho} = \epsilon_\mu m_\rho^2 / f_\rho~,
\end{equation}
where $m_\rho$ is the $\rho$ meson mass and $\epsilon_\mu$ is the
polarization vector.
We have two choices for the vector current $V_\mu$:
the exactly conserved current in eq.(\ref{eq:vcon})
and the four-dimensional local current 
$V_\mu^q(x)={\bar q}(x)\gamma_\mu q(x)$ multiplied
by  the renormalization factor $Z_V$.

In a similar way, the $\pi$ meson decay constant $f_\pi$ is given by 
\be
\vev{0|A_\mu|\pi} = p_\mu f_\pi~,
\ee
where $p_\mu$ denotes the momentum of the pion and 
$A_\mu$ is either the (almost) conserved current
in eq.(\ref{eq:acon})
or the four-dimensional local current 
$A_\mu^q(x)={\bar q}(x)\gamma_\mu\gamma_5 q(x)$ multiplied
by the renormalization factor $Z_A$. 

Perturbative technique in DWQCD is already applied to evaluation of 
the renormalization factors for the bilinear operators\cite{pt_2} and 
three- and four-quark operators\cite{pt_34} consisting of 
four-dimensional quarks in eqs.(\ref{eq:q}) and (\ref{eq:qbar}).
It is shown that $Z_S=Z_P=Z_m^{-1}$ and $Z_V=Z_A$ at $N_s\rightarrow \infty$
for the bilinear operators, which is expected in the case that
the chiral Ward-Takahashi identities hold exactly\cite{pt_2}.
Another desirable feature is 
that the three- and four-quark operators at $N_s\longrightarrow \infty$
can be renormalized without mixing between operators with different 
chiralities, as opposed to the Wilson fermion case\cite{pt_34}.
The peculiar feature in the renormalization of DWQCD, 
however, is an
appearance of the overlap factor $(1-|1-M|^2)Z_w$ for the
four-dimensional quark fields.
The one-loop coefficient $z_w$ of $Z_w$ in \eqn{eqn:zw} is of
$O(10-10^2)$ for 
$|1-M|\simgt 0.1$ without mean-field improvement, which could be
problematic because the present simulation is done for
$M\ge 1.7$.  It is shown in Ref.\cite{pt_2}, however, that
the magnitude of the one-loop coefficient for
$Z_w$ is reduced to $O(1)$ for $|1-{\tilde M}|\simgt 0.1$
after the mean field replacement from $M$ to ${\tilde M}=M+4(u-1)$.
Therefore, comparing the $\rho$($\pi$) meson decay constant 
$f_\rho$($f_\pi$) obtained
from $Z_V V_\mu^q$($Z_A A_\mu^q=Z_V A_\mu^q$) 
with that from $V_\mu^\con$($A_\mu^\con$) gives a good 
testing ground for the validity of the (mean field improved) 
perturbation theory.

Following the notation of Ref.\cite{pt_2}, the renormalization factor
of $V_\mu^q$ is written as
\be
V_\mu^\con=\frac{1}{(1-w_0(M)^2)Z_w(M)}Z_V(M) V_\mu^q,
\ee
where 
\ben
w_0(M)&=&1-M, \\
Z_w(M)&=&1+\frac{g^2}{16\pi^2}C_F z_w(M),
\label{eqn:zw}
\\
Z_V(M)&=&1+\frac{g^2}{16\pi^2}C_F z_V(M),
\een
without mean-field improvement. $C_F=(N^2-1)/(2N)$ is the quadratic
Casimir in SU($N$) gauge group.
In the case of $M=1.819$, we find $1-w_0^2=0.32924$,
$z_w(M)=250.36$ and $z_V(M)=-17.760$ using the results in Ref.\cite{pt_2}.
The large corrections of $1/(1-w_0^2)$ and $Z_w$ could
jeopardize the efficiency of the perturbation theory.
However, once a mean-field improvement is employed,  
the renormalization factors are reexpressed as
\be
V_\mu^\con=\frac{1}{(1-w_0({\tilde M})^2)Z_w^{\mf}({\tilde M})}
u Z_V^{\mf}({\tilde M}) V_\mu^q
\ee
with
\ben
{\tilde M}&=&M+4(u-1), \\
w_0({\tilde M})&=&1-{\tilde M}, \\
Z_w^\mf({\tilde M})&=&1+\frac{g^2}{16\pi^2}C_F z_w^\mf({\tilde M}), 
\label{eq:zwmf}\\
Z_V^\mf({\tilde M})&=&1+\frac{g^2}{16\pi^2}C_F z_V^\mf({\tilde M}).
\label{eq:zvmf}
\een
With the use of $u=0.87781=\sqrt[4]{P}$, where $P$ is 
the expectation value of the plaquette,
we obtain ${\tilde M}=1.3302$, $1-w_0({\tilde M})^2=0.89095$, 
$z_w^\mf(\tilde M)=7.327$ and $z_V^\mf(\tilde M)=-7.3744$ from the results in 
Ref.\cite{pt_2,pt_34}. It is remarkable that the perturbative 
corrections are drastically reduced. 
We determine the coupling constant at the scale $1/a$ in the
$\msbar$ scheme with the aid of the mean-field
improvement,
\be
\frac{1}{g^2_\msbar(1/a)}=P\frac{\beta}{6}-0.13486.
\ee
Incorporating ${g^2_\msbar(1/a)}$ in eqs.(\ref{eq:zwmf}) and (\ref{eq:zvmf}),
we finally obtain the value of the renormalization factor
with the mean-field improvement,
\be
\frac{1}{(1-w_0({\tilde M})^2)Z_w^{\mf}({\tilde M})}
u Z_V^{\mf}({\tilde M})
=0.7504.
\label{eqn:Zfactor}
\ee

If we employ $u=1/(8K_c)$ with $K_c=0.1572$ for the mean field improvement,
we obtain $0.7147$ instead of $0.7504$ in eq.(\ref{eqn:Zfactor}).
Since their difference is smaller than the statistical errors
of the $\rho$ and $\pi$ meson decay constants, we use the value of
eq.(\ref{eqn:Zfactor}) for the renormalization factor in the next subsection.

\subsection{$\rho$ and $\pi$ meson decay constants}

We calculate $f_\rho$ from the ratio of two point functions,
\be
\frac{\sum_{{\vec x},{\vec y}}\vev{
V_\mu^{\con,q}({\vec x},t) J_\rho^\dagger({\vec y},0)}}
{\sum_{{\vec x},{\vec y}}\vev{J_\rho({\vec x},t) 
J_\rho^\dagger({\vec y},0)}}\vev{0|J_\rho|\rho}
\stackrel{t\gg 0}{\longrightarrow}  \vev{0|V_\mu^{\con,q}|\rho},
\ee 
where $J_\rho(x)={\bar q}(x)\gamma_i q(x)$ $(i=1,2,3)$ is 
the interpolating field for the $\rho$ meson. The amplitude
$\vev{0|J_\rho|\rho}$ is obtained by fitting the two-point function with
\be
\sum_{{\vec x},{\vec y}}\vev{J_\rho({\vec x},t) J_\rho^\dagger({\vec y},0)}
=V_s\frac{|\vev{0|J_\rho|\rho}|^2}{2m_\rho}
\left({\rm e}^{-m_\rho t}+{\rm e}^{-m_\rho(32-t)}\right), 
\ee
where the fitting range is chosen to be $8\le t\le 24$. 

\begin{figure}[t]
\hspace{10pt}\psfig{file=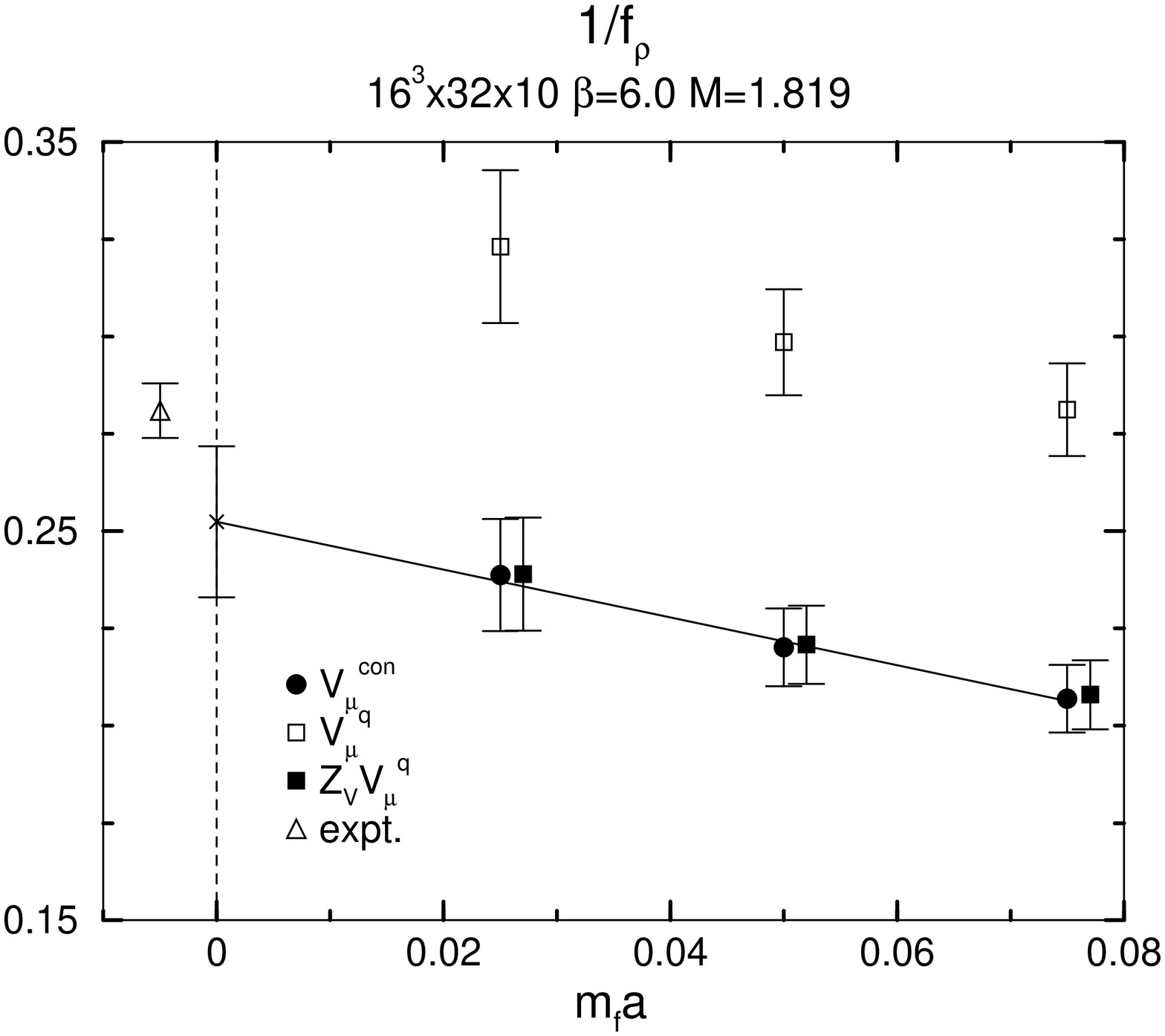,width=100mm,angle=-90}
\caption{$f_\rho$ as a function of $m_f a$
at $M=1.819$ and  $N_s=10$ together with  the experimental value.
Data points for $Z_V V_\mu^q$ (filled squares) are slightly
displaced horizontally for clarity. 
Solid line shows linear extrapolation of the results 
for the conserved current.
}
  \label{fig:FV}
\end{figure}

In Fig.\ref{fig:FV} we plot 
the $\rho$ meson decay constants obtained
from $V_\mu^q$ and $V_\mu^\con$ as a function of $m_f$.
We observe that both results are consistent once the perturbative corrections
are applied to $V_\mu^q$. This is an encouraging evidence to 
show the validity of the perturbation theory with the
mean-field improvement. Still, the scaling violation 
effects should be checked. 
The value at the chiral limit $m_f=0$,
which is obtained by extrapolating the results for the conserved current
linearly, is found to be slightly smaller than
the experimental value.

Let us turn to the $\pi$ meson decay constant $f_\pi$.
We obtain $f_\pi$ from correlation functions of the axial vector current
$A_\mu(x)$ and the interpolating field for the $\pi$ meson
$J_\pi(x)={\bar q}(x)\gamma_5 q(x)$,
\be
\sum_{{\vec x},{\vec y}}\vev{X({\vec x},t) Y^\dagger({\vec y},0)}
=V_s\frac{\vev{0|X|\pi}\vev{\pi|Y^\dagger|0}}{2m_\pi}
\left({\rm e}^{-m_\pi t}+{\rm e}^{-m_\pi(32-t)}\right), 
\ee
where $X,Y = J_\pi, A_4$.
The fitting range is chosen to be $9\le t\le 23$. 

\begin{figure}[t]
\hspace{10pt}\psfig{file=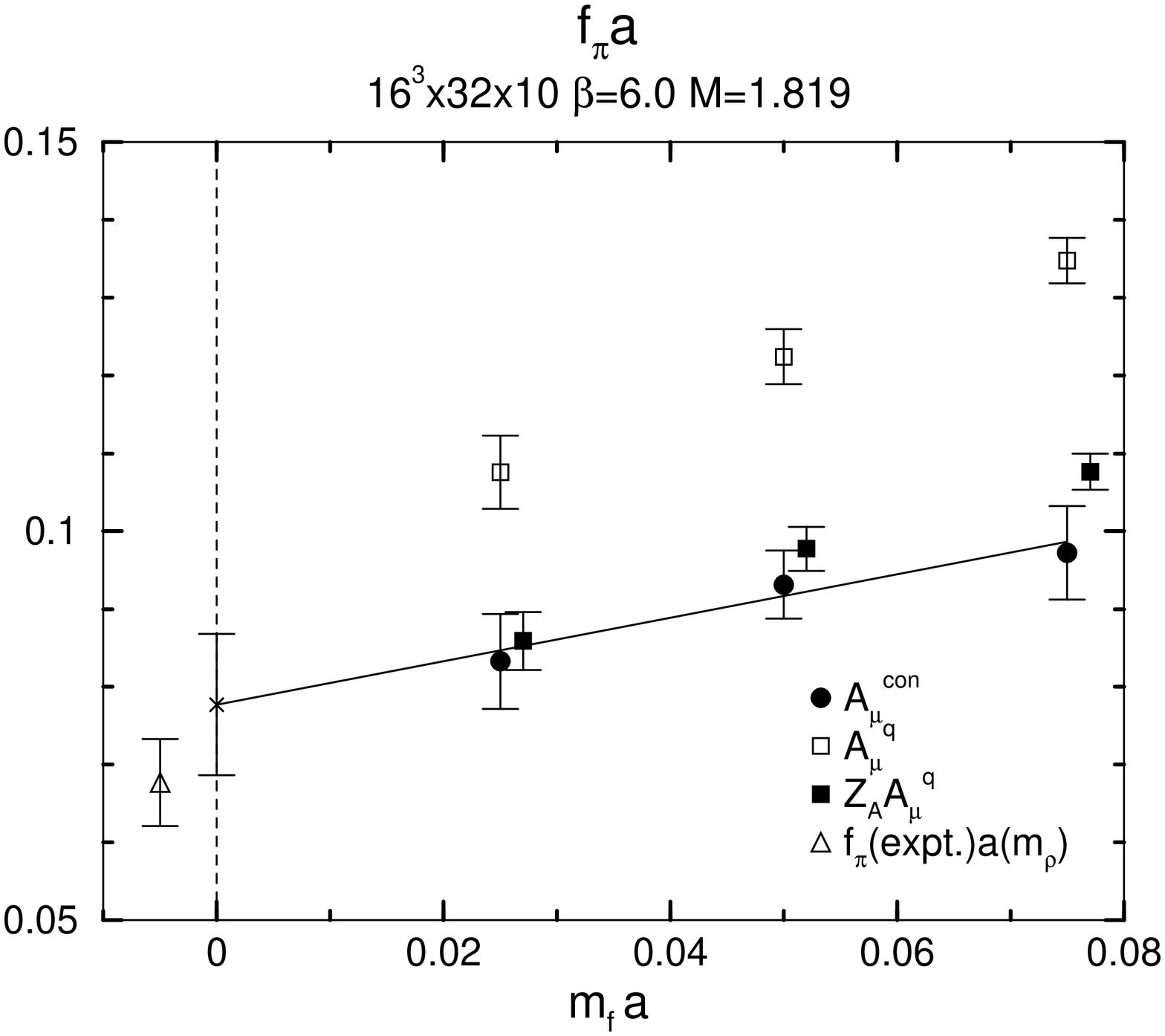,width=100mm,angle=-90}
\caption{$f_\pi$ in lattice unit as a function of $m_f a$
at $M=1.819$ and  $N_s=10$ together with  the experimental value. 
Data points for $Z_A A_\mu^q$ (filled squares) are slightly
displaced horizontally for clarity. 
Solid line shows linear extrapolation of the results 
for the conserved current.
}
  \label{fig:FpiSRnmPlaq_M=1.819}
\end{figure}

\begin{figure}[t]
\hspace{10pt}\psfig{file=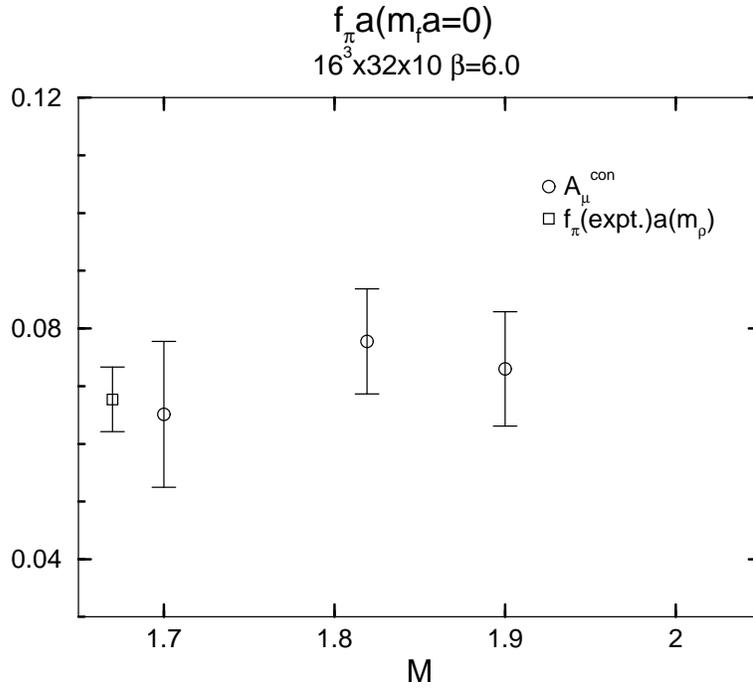,width=100mm,angle=-90}
\caption{$f_\pi$ in lattice unit extrapolated to $m_f a \to 0$ 
as a function of $M$ at  $N_s=10$ together with  the experimental value. 
}
  \label{fig:Fpi_M}
\end{figure}

Figure~\ref{fig:FpiSRnmPlaq_M=1.819} shows a comparison of  
the $\pi$ meson decay constants obtained
from $A_4^q$ and $A_4^\con$. 
We find the same situation as in the case of the
$\rho$ meson decay constant: the perturbative corrections
sufficiently compensate the difference between $A_4^q$ and $A_4^\con$.
Linear extrapolation of the results for the conserved current 
gives the value at the chiral limit $m_f=0$, which is
roughly consistent with the experimental value.
We also show the $M$ dependence of $f_\pi$ at $m_f=0$ 
in Fig. \ref{fig:Fpi_M}.
Although the results slightly depend on the choice of $M$,
the differences are smaller than current statistical errors.

\section{Conclusions and discussions}
\label{sec:conclusions}

We have investigated the chiral properties of DWQCD 
by measuring the pion mass and the explicit chiral symmetry
breaking term in the PCAC relation.
Their $N_s$ dependence up to $N_s=10$ seem to be consistent
with exponential decay in $N_s$,
which indicates that the chiral symmetry 
is restored at $N_s\to\infty$ limit, as expected.
More extensive study with larger $N_s$, however, is required
to confirm this conclusion.

We also calculate the pion mass at negative $m_f$ 
to examine whether there exists the parity broken phase in DWQCD. 
Our result is consistent with the existence of the parity broken phase.

The validity of the perturbation theory in DWQCD is
tested by calculating the $\rho$ and $\pi$ meson decay constants
from the four-dimensional local currents and the conserved ones.
We find that the difference between both currents is 
sufficiently compensated with the perturbative renormalization factor 
up to the one-loop level with the aid of the mean-field improvement.
This is an encouraging result, though the magnitude of the scaling violation 
on these quantities should be checked further.

\section*{Acknowledgement}
Numerical calculations for the present work have been carried out 
on VPP500/30 at Science Information Processing Center at University of
Tsukuba.
This work is supported in part by the Grants-in-Aid of the Ministry of
Education(No. 2373).
T.I., Y.K. and Y.T. are JSPS Research Fellows.



\begin{table}[h]
\begin{center}
\caption{\label{tab:parameter} Simulation parameters
on $16^3\times 32\times N_s$ lattices in the quenched QCD. } 
\begin{tabular}{llll}
$N_s$ & $4$ & $6$ & $10$ \\
\hline
No.~conf. & $20$               & $10$    & $30(M=1.819), 20$    \\   
$M$       & $1.7$              & $1.819$ & $1.7$   \\    
          & $1.75$             &         & $1.819$ \\    
          & $1.819$            &         & $1.9$   \\    
          & $1.85$             &         &         \\
          & $1.9$              &         &         \\
          & $1.95$             &         &         \\
          & $2.0$              &         &         \\    
$m_fa$    & $0.075$            & $0.075$ & $0.075$ \\   
          & $0.050$            & $0.050$ & $0.050$ \\   
          & $0.025$            & $0.025$ & $0.025$ \\   
          & $-0.08\;(M=1.819)$ &         &         \\   
          & $-0.086\;(M=1.819)$ &         &         \\   
          & $-0.093\;(M=1.819)$ &         &         \\   
          & $-0.10\;(M=1.819)$ &         &         \\   
          & $-0.12\;(M=1.819)$ &         &         \\   
\end{tabular}
\end{center}
\end{table}

\end{document}